\documentclass[fleqn,twoside]{article}
\usepackage{espcrc2}

\usepackage{graphicx}
\usepackage[figuresright]{rotating}

\newcommand{\AmS}{{\protect\the\textfont2
  A\kern-.1667em\lower.5ex\hbox{M}\kern-.125emS}}

\def\chpt{\mbox{\raise.4ex\hbox{$\chi$}PT}}
\def\schpt{\mbox{S\raise.4ex\hbox{$\chi$}PT}}

\def\Figref#1{Figure~\ref{fig:#1}}
\def\half{{\scriptstyle \raise.2ex\hbox{${1\over2}$}}}
\def\fourth{{\scriptstyle \raise.2ex\hbox{${1\over4}$}}}

\def\gtwid{\raise.3ex\hbox{$>$\kern-.75em\lower1ex\hbox{$\sim$}}}
\def\ltwid{\raise.3ex\hbox{$<$\kern-.75em\lower1ex\hbox{$\sim$}}}

\def\eg{{\it e.g.},\ }
\def\et{{\it et al.}}

\def\cO{{\cal O}}

\def\prd#1{Phys.\ Rev.\ {\bf D#1}}

\def\berlin{Nucl.\ Phys.\ {\bf B} (Proc.\ Suppl.) {\bf 106-107} (2002)}
\def\boston{Nucl.\ Phys.\ {\bf B} (Proc.\ Suppl.) {\bf 119} (2003)}
\def\tsukubathree{these proceedings}

\def\MeV{{\rm Me\!V}}
\def\GeV{{\rm Ge\!V}}

\newcommand{\BE}{\begin{equation}}
\def\EE{\end{equation}}
\def\BEA{\begin{eqnarray}}
\def\EEA{\end{eqnarray}}

\title{
\null\vskip -1cm
Pion and kaon physics with improved staggered 
quarks\thanks{presented by C.\ Bernard at {\it Lattice 2003}, Tsukuba,
Japan, July 15-19, 2003}
}

\author{C.~Aubin,\hskip-0.03in
\address{{\vskip-0.10in{\hskip 0.07in Department of Physics, Washington
University, St.~Louis, MO 63130, USA \vskip -.1truecm}}} 
C.~Bernard,$\null^{\rm a}$ 
\hskip-0.03in
C.~DeTar,\hskip-0.03in
\address{Physics Department, University of Utah, Salt Lake City, UT 84112, USA \vskip -.1truecm} 
Steven~Gottlieb,\hskip-0.03in
\address{Department of Physics, Indiana University, Bloomington, IN 47405, USA \vskip -.1truecm} 
E.B.\ Gregory,\hskip-0.03in
\address{Department of Physics, University of Arizona, Tucson, AZ 85721, USA \vskip -.1truecm} 
Urs~M.~Heller,\hskip-0.03in
\address{American Physical Society, One Research Road, Box 9000, Ridge, NY 11961, USA \vskip -.1truecm} 
J.E.~Hetrick,\hskip-0.03in
\address{Physics Department, University of the Pacific, Stockton, CA 95211, USA \vskip -.1truecm} 
J.\ Osborn,$\null^{\rm b}$ 
R.~Sugar\hskip0.005in
\address{Department of Physics, University of California, Santa Barbara, CA
93106, USA \vskip -.1truecm} 
and
D.~Toussaint$\null^{\rm d}$
\advance\baselineskip -2pt
} 
       
\begin{document}

\begin{abstract}
We compute pseudoscalar meson masses and decay constants using staggered
quarks on lattices with three flavors of sea quarks
and lattice spacings $\approx\!0.12\;$fm and $\approx\!0.09\;$fm. 
We fit partially quenched results to
``staggered chiral perturbation
theory'' formulae, thereby taking into account 
the effects of taste-symmetry violations. Chiral logarithms
are observed.
From the fits we calculate $f_\pi$ and $f_K$,
extract Gasser-Leutwyler parameters of the
chiral Lagrangian, and (modulo rather large perturbative 
errors) find the light and strange quark masses.
\vspace{-.2truecm}
\end{abstract}

\maketitle
The masses and decay constants of light pseudoscalars can be computed with high
precision at fixed lattice spacing and quark mass.  
Assuming that the chiral and continuum extrapolations can be performed with controlled
systematic errors,
such a calculation therefore can provide an accurate
determination of the physical properties of the $\pi$-$K$ system, as well as a
sensitive check on our algorithms and methods.

Our simulations have
three light dynamical flavors with
the Asqtad action\cite{ASQTAD}, with $\cO(\alpha_s a^2)$ errors in general, 
but $\cO(\alpha^2_s a^2)$ taste-violations, which require 2-gluon exchange.
One group of lattices (``coarse'')
has $a\! \approx\! 0.12\;$fm, the dynamical strange quark mass held fixed at 
a nominal value 
$am_s =0.05 $,  and five different values of dynamical light
quark mass $m_u\!=\!m_d\!\equiv\! m_q$. The second group (``fine'')
has $a\!\approx\! 0.09\;$fm, a nominal value $am_s=0.031$, and two values of
$m_q$. The nominal strange quark mass turned out to be about 18\% (10\%)
larger than the physical mass, $m_s^{\rm phys}$, on the
coarse (fine) lattices.
Table \ref{tab:lattices} shows the lattice parameters used.

\begin{table}
\begin{center}
\begin{tabular}{|c|c|c|c|}
\hline
$am_q$ / $am_s$  & \hspace{-1.0mm}$10/g^2$ & size & number \\
\hline
0.03  / 0.05   & 6.81 & $20^3\times64$ & 262 \\
0.02  / 0.05   & 6.79 & $20^3\times64$ & 485 \\
0.01  / 0.05   & 6.76 & $20^3\times64$ & 608 \\
0.007  / 0.05   & 6.76 & $20^3\times64$ & 447 \\
0.005  / 0.05   & 6.76 & $24^3\times64$ & 137 \\
\hline
\hline
0.0124  / 0.031   & 7.11 & $28^3\times96$ & 531 \\
0.0062  / 0.031   & 7.09 & $28^3\times96$ & 583 \\
\hline
\end{tabular}
\end{center}
\caption{Lattice parameters.
The lattice sets above the double line are ``coarse;'' those below are ``fine.''
}
\label{tab:lattices}
\vskip -.8truecm
\end{table}
    
The relative scale among coarse (or, separately, fine) lattices is kept fixed
using the length $r_1$ \cite{SOMMER,MILC_POTENTIAL} from the static quark potential.
We reduce statistical fluctuations in $r_1/a$ by using a ``smoothed'' $r_1/a$ 
coming from a 4-parameter fit to a smooth function of $10/g^2$ and  
$m_{\rm tot}\equiv 2m_q + m_s$. 
The absolute scale is taken from $\Upsilon$ $2S$-$1S$ or $1P$-$1S$ splittings,
determined by the HPQCD group \cite{LEPAGE_PC}
on the lattices with $m_q=0.2m_s$.
We extrapolate to the continuum linearly in $\alpha_s a^2$ and get
$r_1 = 0.317(7)\;$fm.  
Since the physical $r_1$ has some (small) 
dependence on sea quark masses, fixing $a$ using $r_1/a$ can introduce
some spurious dependence on mass \cite{SOMMER_LAT03}.  This
has a negligible effect on quantities like $f_\pi$,
but could be a significant systematic for the Gasser-Leutwyler parameters ($L_i$).  
Estimates of this effect are in progress.

On the coarse (fine) lattices, quark propagators are computed every 6 time units
for 9 (8) values of valence mass between $0.1m_s$ ($0.14m_s$) and $m_s$. 
We determine meson masses and decay constants from simultaneous
fits to wall-point and point-point correlators.
The complete covariance matrix of decay constants and masses 
on each lattice set is computed using single elimination jackknife;
we then adjust for autocorrelations by scaling the matrix by a factor
estimated by blocking lattices in groups of four.

We first fit
squared masses of various-taste pions to the tree-level form \cite{LEE_SHARPE}:
linear in quark mass plus a constant splitting (at fixed $a$) for each
taste multiplet.
Although the fits are poor, they
do give the masses within $\sim\! 5\%$. The slopes and
splittings determined may then be consistently used as inputs to the
one-loop terms in the chiral log fits below. The ratio of splittings on
the fine lattices to those on the coarse lattices is $\approx\!0.35$.
This is consistent with expectation that taste violations go like
$\cO(\alpha^2_s a^2)$: using
$\alpha_s = \alpha_V(q^*)$ at one-loop \cite{DAVIES_LAT02} and $q^*=3.33/a$,
this ratio is 0.375.

We then compare the (partially quenched) decay constants and 
squared meson masses 
to one-loop chiral log forms (including finite 
volume corrections) computed in
staggered chiral perturbation theory (\schpt) \cite{CA_CB}. 
Up through NLO, we have 10 free parameters in the chiral expressions.
Of these, two appeared previously at tree level: 
the decay constant $f$ and the ratio, $\mu$, of squared Goldstone-meson mass 
to the sum of valence quark masses,
$m_x+m_y$.  There are also analytic contributions proportional to
$L_4$, $L_5$, $2L_8-L_5$, and $2L_6-L_4$.  Finally there are four $\cO(a^2)$
parameters that appear because of taste violation \cite{CA_CB}: $a^2\delta'_A$ and $a^2\delta'_V$,
which are the taste-violating ``hairpin'' parameters that
enter into NLO chiral logs, and $a^2C$ and $a^2F$, which give analytic,
$\cO(ma^2)$ contributions.

Since the data is very precise  ($\sim\!0.1$--$0.4\%$ statistical errors)
and we include quark masses
as large as $m_s^{\rm phys}$ (to extract $K$ physics), 
we must go beyond NLO to get
good fits.  We include all $\cO(m^3)$ NNLO analytic
terms.
NNLO chiral logs and taste-violating analytic terms are 
unknown and not included.  But for larger masses, where NNLO
terms are non-negligible, the logarithms should be changing slowly and thus
well represented by analytic terms.  Similarly, for larger masses,
$\cO(m^3)$ terms should be more important than taste-violating
$\cO(m^2a^2)$ or $\cO(ma^4)$ terms. 
When expressed in ``natural'' chiral units, the 10 NNLO $\cO(m^3)$ parameters (5 for the decay
constant and 5 for the mass) should have coefficients of $\cO(1)$
if chiral perturbation theory (\chpt) is well behaved. 
We constrain them to have standard deviation $\sigma=1$ around 0 using Baysean priors \cite{LEPAGE_BAYSE}.

The strength of the chiral log terms is governed by
the ``chiral coupling,'' $1/(16\pi^2\tilde f^2)$. Typically, one takes $\tilde f=f$,
where $f$ is the bare (tree-level) decay constant.  For better convergence of \chpt, 
it seems reasonable to put a {\it physical} parameter here, \eg, $\tilde f = f_\pi$ or 
$\tilde f=f_K$ (the difference is NNLO).  
In practice, we try 3 approaches: (1) $\tilde f= f$, (2) $\tilde f= f_\pi$, and (3)
$\tilde f = f_\pi/\sqrt{\omega}$, where $\omega$ is a new fit parameter allowed to vary
around 1 with  $\sigma=0.1$.
So far, the best fits have been obtained with choice (3); we currently use
such fits for central values.  Choices (1) and (2) are now also giving 
acceptable fits. Only choice (2) has been included in systematic error 
estimates below, but choice (1) will be added in the future.

To test convergence of \chpt, a NNNLO term in valence quark
mass is included for both decay constants and masses.  
Such terms are found to have small coefficients; 
we plan to eliminate them 
for the fits that determine central values.

We fit both coarse and fine lattice data simultaneously.
The 4 taste-violating parameters are forced to change by the factor
of 0.375 (ratio of $\alpha_s^2 a^2$) in going from coarse to fine.  
The remaining physical parameters are expected to 
differ between coarse and fine lattices by
$\alpha_s a^2 \Lambda^2_{QCD}\sim 2\%$.  
We therefore also include an additional ``scaling'' parameter
for each physical parameter.  The scaling parameter is the fractional difference between 
the physical parameter on the
coarse and fine lattices. The scaling parameters are constrained to
be 0 with $\sigma$ of 0.02 or 0.025 in central value fits;  
this is changed to 0.01 or 0.04 in fits used to estimate systematics.

Finally, the 4 parameters used in smoothing $r_1/a$ are allowed to vary by 1 $\sigma$.
The total is 46 parameters (or 44 with choices (1) or (2) above), although many are 
tightly constrained.

\begin{figure}[tb]
\null
\vspace{-.2truecm}
\includegraphics[width=2.9truein]{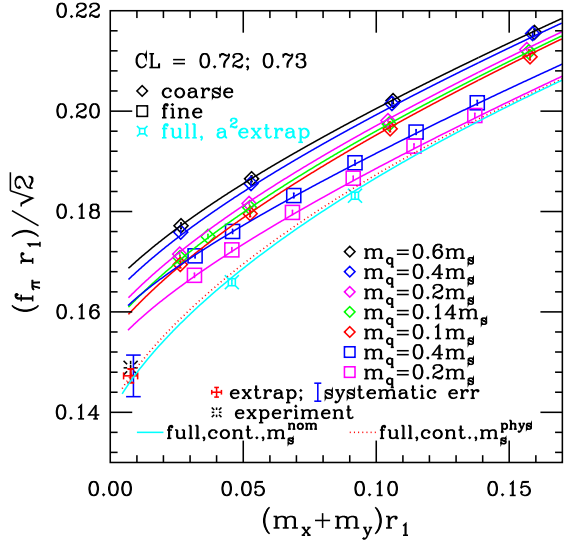}
\vspace{-1.0truecm}
\caption{
Decay constants with $m_x\!=\!m_y$. 
The two
cyan 
``fancy squares'' 
are full QCD ($m_x\!=\!m_y\!=\!m_q$) points, continuum extrapolated
at fixed quark mass.  The lowest two lines 
(dotted red and cyan)
use continuum-extrapolated fit parameters,
$m_x\!=\!m_y\!=\!m_q$, and either the physical or the (fine lattice) nominal
strange mass.
The first confidence level (CL) is computed in the standard way; 
the second treats the Baysean priors as if they were additional data points.
}
\label{fig:fpi} 
\end{figure}

\Figref{fpi} shows some decay constants from our central-value fit
to the partially quenched data with $m_x,m_y \ltwid 0.8 m_s^{\rm phys}$ (194 points).  
We determine physical quark masses from the
meson-mass part of this fit (see Ref.~\cite{GOTTLIEB_LAT03})
with $m_\pi$ and $m_K$ as input;
the decay constants are then extrapolated to the physical masses and to the
continuum.

Our preliminary results for decay constants are:
\vskip -.5truecm
\begin{eqnarray}\label{eq:f_results}
f_\pi & = &  129.3 \pm 1.1 \pm 3.5 \; \MeV \nonumber\\
f_K & = &  155.0 \pm 1.8 \pm 3.7 \; \MeV \nonumber\\
f_K/f_\pi  & = & 1.201(8)(15)\ ,
\end{eqnarray}
where the first error is statistical; the second, systematic.
The largest error on $f_\pi$ and $f_K$ 
is the 2.2\% scale uncertainty. A chiral and continuum extrapolation error of
$\approx\!1.5\%$ has been estimated by 
considering alternative fits.
The results agree with experiment within 
errors.

Preliminary results (in units of $10^{-3}$) for  the $L_i$ at chiral scale $m_\eta$ are:
\vskip -.5truecm
\begin{eqnarray}\label{eq:Li_results}
2L_8\! -\! L_5 \!=\! -0.1(1)({}^{+1}_{-3})\;; &&
\!\!L_5 = 1.9(3)({}^{+6}_{-2}) \nonumber\\
2L_6\! - \!L_4 \!=\! 0.5(2)({}^{+1}_{-3})\;; &&
\!\!L_4 = 0.3(3)({}^{+7}_{-2})\;.
\end{eqnarray}
The errors here are dominated by differences over
The result for $2 L_8\! -\!L_5$ is well outside the range that allows for
$m_u=0$ \cite{MUZERO}.

Finally, our preliminary results for quark masses at scale $2\,\GeV$ are
$m_s^{\overline{MS}} =  70(15)\; \MeV$ and $\hat m^{\overline{MS}} = 2.7(6)\; \MeV$, 
where $\hat m$ is the average of the $u$ and $d$ masses.
These results use the 1-loop perturbative result for mass renormalization \cite{ZM1,ZM2}.
The rather large error is dominated by the $\cO(\alpha_s^2)$ correction to the 
renormalization constant, estimated in Ref.~\cite{ZM1} to be $\sim\!20\%$.

Additional discussion of the calculation and
results appears in Ref.~\cite{GOTTLIEB_LAT03}.


\begin{thebibliography}{9}
\bibitem{ASQTAD}
See, \eg 
C.~Bernard \et,       
Phys. Rev. D {\bf 61} (2000) 111502 and references therein.

\bibitem{SOMMER}
R. Sommer, Nucl. Phys. {\bf B411} (1994) 839.

\bibitem{MILC_POTENTIAL}
C.~Bernard {\it et al.},
Phys.\ Rev.\ D {\bf 62} (2000) 034503.

\bibitem{SOMMER_LAT03}
See, \eg R. Sommer, \tsukubathree.


\bibitem{LEPAGE_PC}
G.P.\ Lepage, private communication.

\bibitem{LEE_SHARPE}
W.\ Lee and S.\ Sharpe,
\prd{60} (1999) 114503.


\bibitem{DAVIES_LAT02}
C.\ Davies \et,
\boston\ 595. 


\bibitem{CA_CB}
C.\ Aubin and C. Bernard,
\prd{68} (2003) 034014, 
hep-lat/0306026, 
and hep-lat/0308036 (\tsukubathree); 
C.\ Bernard, \prd{65} (2001) 054031.

\bibitem{LEPAGE_BAYSE}
See G.P.~Lepage \et,
\berlin\ 12.

\bibitem{GOTTLIEB_LAT03}
S.\ Gottlieb, plenary talk, \tsukubathree.

\bibitem{MUZERO}
D.~Kaplan and A.~Manohar,
Phys.\ Rev.\ Lett.\  {\bf 56} (1986) 2004;
A.~Cohen, D.~Kaplan and A.~Nelson,
JHEP {\bf 9911} (1999) 027.

\bibitem{ZM1}
J.~Hein, {\it et al.},
\berlin\ 236;
{\bf 119} (2003) 317.

\bibitem{ZM2}
T.~Becher and K.~Melnikov,
Phys.\ Rev.\ D {\bf 68} (2003) 014506.

      \end{thebibliography}
\end{document}